# Title: Not your private tête-à-tête: leveraging the power of higher-order networks to study animal communication

*(Special issue: 'The power of sound – unravelling how vocal communication shapes group dynamics')*

Iacopo Iacopini[1], Jennifer R Foote[2], Nina H Fefferman[3,4,5], Elizabeth P Derryberry[3], Matthew J Silk[6,7]*

[1]Network Science Institute, Northeastern University London, London, E1W 1LP, United Kingdom
[2]Department of Biology, Algoma University, Sault Ste. Marie, ON, Canada
[3]Department of Ecology and Evolutionary Biology, University of Tennessee, Knoxville, TN, USA
[4]Department of Mathematics, University of Tennessee, Knoxville, TN, USA
[5]NIMBioS, University of Tennessee, Knoxville, TN, USA
[6]CEFE, Univ Montpellier, CNRS, EPHE, IRD, Montpellier, France
[7]Institute of Ecology and Evolution, University of Edinburgh, Edinburgh, UK

*corresponding author: matthewsilk@outlook.com

ORCIDs
II – https://orcid.org/0000-0001-8794-6410
JF - https://orcid.org/0000-0001-9128-3496
NF – https://orcid.org/0000-0003-0233-1404
ED – https://orcid.org/0000-0002-8248-9748
MS – https://orcid.org/0000-0002-8318-5383

## Abstract

Animal communication is frequently studied with conventional network representations that link pairs of individuals who interact, for example, through vocalisation. However, acoustic signals often have multiple simultaneous receivers, or receivers integrate information from multiple signallers, meaning these interactions are not dyadic. Additionally, non-dyadic social structures often shape an individual's behavioural response to vocal communication. Recently, major advances have been made in the study of these non-dyadic, higher-order networks (e.g., hypergraphs and simplicial complexes). Here, we show how these approaches can provide new insights into vocal communication through three case studies that illustrate how higher-order network models can: a) alter predictions made about the outcome of vocally-coordinated group departures; b) generate different patterns of song synchronisation than models that only include dyadic interactions; and c) inform models of cultural evolution of vocal communication. Together, our examples highlight the potential power of higher-order networks to study animal vocal communication. We then build on our case studies to identify key challenges in applying higher-order network approaches in this context and outline important research questions these techniques could help answer.

# 1. Introduction

Quantifying the role of communication in the social coordination of animals has long been a topic of considerable interest in ecology and evolution (1,2). Network analysis is a useful tool to study patterns of communication within animal groups (3,4) and populations (5). For example, vocal communication may be used to maintain close social bonds (3) or play a key role in wider group coordination by enabling individuals to maintain weak social connections without interacting closely (4). However, while it is widely acknowledged that vocal communication frequently involves more than two individuals within each "interaction" (2), most existing analyses have only used dyadic representations of these communication networks.

Dyadic representations can capture valuable aspects of animal communication network structure (1,2). However, there are often social mechanisms acting on interactions of three or more individuals at a time. Eavesdropping (6–8) and audience effects (9,10) both represent examples of when non-dyadic animal communication shapes ecological and evolutionary outcomes. Accounting for the multibody nature of such interactions prevents losing relevant information. Powerful new higher-order network approaches (11–14) encode these non-pairwise interactions between agents, helping us quantify the importance of multibody interactions in driving group dynamics and wider social coordination. By explicitly representing multibody interactions, higher-order approaches capture the rich set of dynamics introduced by including the non-dyadic components of communication networks.

Vocal communication involves the transmission of information encoded in sound from one (or more) signallers to one or more receivers (1,15). Individuals integrate information received to inform their social decision-making (e.g. (16–18)). Consequently, tools to understand social transmission from network science can help understand the outcomes of vocal communication. Typically, social transmission occurs as a *complex contagion* (19), in which the probability of an individual changing state (i.e. altering its behaviour in the light of the information it receives) is not an additive function of the state of its neighbours (i.e., each interaction with a signaller is associated with an independent probability of behaviour change). These social effects on behaviour spread are not limited to humans. For example, non-human animals can change their state when most of their social connections behave in a particular way (conformity; (19)). For instance, complex contagion models best reproduce the outcome of behavioural transmission experiments in fish schools (20). While it is possible to model the dynamics of these complex contagions across dyadic networks, many forms of behavioural spread are inherently mediated by multibody interactions, naturally calling for higher-order network approaches (21).

Here we introduce higher-order network approaches such as hypergraphs, simplicial sets, and simplicial complexes (12,13). We illustrate how they can be applied to study animal communication using a series of toy models based on real-world case studies across a range of temporal scales. We then provide an overview of the tools available to empiricists keen to explore the higher-order structure of their data. Finally, we summarise potential future directions for research in this area, highlighting possible synergies between research advances for both the animal behaviour and network science communities.

## 2. Structure of higher-order communication networks

Higher-order network approaches make it possible to represent multibody interactions that involve two or more individuals at a time (Fig. 1). There are three common higher-order network representations applicable to vocal communication networks: hypergraphs (Fig. 1a), simplicial sets, and simplicial complexes (Fig. 1b) with overviews of these approaches provided elsewhere (12,13,22,23).

Hypergraphs extend dyadic networks to enable edges (termed hyperedges) among any number of nodes (Fig. 1a, Fig. 1c). Hypergraphs can still represent dyadic interactions between pairs of individuals but also capture situations in which communication occurs among any number of individuals as one hyperedge, rather than multiple dyadic edges (Fig. 1c). This is important as vocal communication dynamics may differ between situations with multiple signallers and/or receivers (whether intended or not) and those in which communication is only dyadic (5,8–10,24). The connectivity of a hypergraph can be encoded as an incidence matrix; an explicitly higher-order representation (Fig. 1e) that link individuals (nodes) to specific group interactions (hyperedges), akin to a group-by-individual matrix in animal social behaviour research (25). Extensions to hypergraph approaches allows hyperedges to be directed — potentially weighted— hyperarcs (26), which may be useful to study some communication networks.

Simplicial sets (not illustrated) are broadly equivalent to hyperedges and represent an alternative mathematical framework to represent higher-order interactions (27). For example, a 0-simplex is the same as a node in a network, a 1-simplex the same as an edge, a 2-simplex an interaction involving three individuals, and so on. Unlike hypergraphs, one can represent the influence of interactions without the presence of the constituent individuals in a simplicial set representation (24,28). However, this extension will only occasionally be useful in studying communication networks, so we do not focus on it here.

We describe simplicial sets to introduce simplicial complexes (Fig. 1f-h). A simplicial complex is a specific type of simplicial set, which must contain all nested lower-order simplices, i.e., requires downward closure. For example, a simplicial complex that contains the simplex $(i, j, k)$ must also contain the simplices $(i, j)$, $(i, k)$, $(j, k)$, $(i)$, $(j)$ and $(k)$. This extra requirement makes simplicial complexes somewhat limited in their ability to faithfully represent complex systems, as at times the inclusion of all the possible sub-interactions would result in a tight constraint. Nevertheless, working with simplicial complexes brings a lot of mathematical advantages, as it allows for the use of tools from *topological data analysis* (TDA). Readers interested in exploiting the recent advances in TDA to study higher-order communication landscapes can read (29).

One directly applicable approach for communication networks is to construct a simplicial complex of a random geometric hypergraph (30) based on individual locations and their audible radii. Constructing this representation is the same approach as (5) but, instead of constructing a dyadic network, builds a simplicial complex representing the potential for higher-order interactions. Similar higher-order structures that could be used are the Vietoris-Rips (Fig. 1h) and Čech complexes (31). The Vietoris-Rips complex adds higher-order simplices to cliques in a dyadic network, i.e. if the 1-simplices $(i, j)$, $(i, k)$ and $(j, k)$ all exist then the 2-simplex $(i, j, k)$ will be added. The Čech complex corresponds to the distribution of 0-simplices in space, defining an interaction radius and adding simplices corresponding to the intersection of the circles defined by each radius. This could be used to construct the information landscape for receivers navigating through a signalling collective, such as females listening to a chorus of simultaneously vocalising males (e.g., (32)).

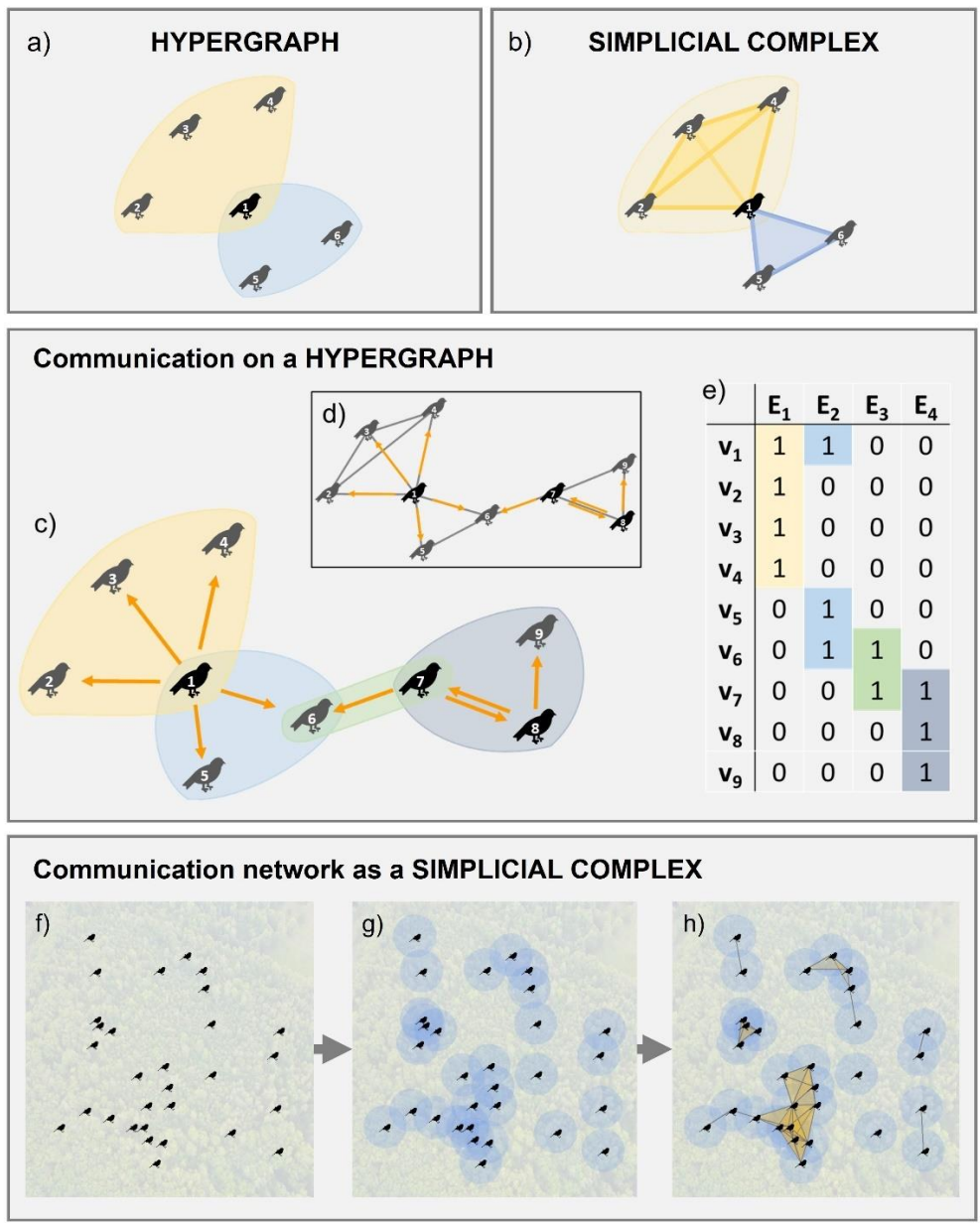

*Figure 1. An introduction to higher-order networks for animal communication based on a) hypergraphs (network edges can connect any number of individuals) and b) simplicial*

*complexes (all nested lower-order interactions must be included in the network object). We then show examples of higher-order networks applied to animal communication. c) Communication as a dynamical process on a series of higher-order social interactions (indicated by the coloured hyperedges). Exploiting higher-order interactions in this way explicitly quantifies biological phenomena such as audience effects and eavesdropping effects that cannot be fully captured using d) the dyadic network representation of the same system. e) Hypergraphs are easily represented as incidence matrices that link individuals to social groupings or events (equivalent to group-by-individual matrices) or can be stored as lists of the events. Communication networks can often be well represented using simplicial complexes, such as that produced in the process illustrated by f-h): f) A researcher locates the position of all individuals within a population; g) using data on the audible radius of different individuals (here assuming no individual variation for illustrative purposes) they can then calculate different types of simplicial complexes to represent the higher-order communication network; h) we illustrate the first-order (grey lines) and second-order simplices (semi-transparent yellow polygons) of the Vietoris-Rips complex for this example.*

## 3. Applications of higher-order communication networks

An important consideration when applying these approaches is whether the higher-order structure is a) an aspect of the social structure over which communication occurs (Fig. 1a), or b) an integral part of the communication process itself (Fig. 1f). An example of the former case would be how individuals produce and respond to contact calls. In this case the contact call is a directed signal on a higher-order social network. The underlying social structure might influence whether the signaller produces a contact call or the response of receivers to the signal. For instance, a receiver may respond differently when in a dyad with the signaller versus when part of a larger group. Consequently, incorporating higher-order social structure can alter predictions about how vocal communication spreads information through a group. An example of the latter case would be a chorus. The vocal communication itself can be encoded into a higher-order network, as the information available is altered by whether a receiver can hear one, two, or more signallers. Our first case study is an example of modelling vocal communication as a directed signal on a higher-order social structure. We then provide two examples that treat the structure of the communication network itself as higher-order; the first incorporates higher-order structure to vocal communication networks to model signal synchronisation, and the second explores how higher-order structure can shape long-term patterns of group coordination and culture.

### 3.1 Case Study 1: Group coordination and consensus decision-making

In many species that rest or feed in groups, group departures are coordinated using vocal communication (33,34). Frequently, quorum decision-making allows group departures to be fully or partially coordinated (33,35). For example, in western jackdaws *Corvus monedula* call intensity increases immediately prior to a group departure and experimental playback leads to earlier departures from communal roosts (33). Similarly, red-fronted lemurs *Eulemur rufifrons* also increase call frequency prior to collective departures enabling group coordination (34).

We can model these behavioural states as contagions on social networks to quantify transitions from individuals a) not calling to calling and b) being present in the group to departed. Because social contagions are often best considered complex contagions and animal groups frequently contain higher-order social structures such as subgroups (11) or family units (36), these systems are suited to modelling as a directed, dyadic signal (vocal communication) across a higher-order network structure. Incorporating this higher-order social structure could make meaningful differences to the predictions made about group coordination, helping to elucidate how partial- and full-consensus decisions are reached.

Here we provide an example of vocally-coordinated departures from a group containing higher-order social structure. Our example is inspired by foraging and roosting flocks of light-bellied brent geese *Branta bernicla hrota*. In the non-breeding season, this species feeds and roosts in fission-fusion social groups (Fig. 2a). Groups contain multiple family units as juveniles accompany adult individuals for their first winter. As in other goose species, group departures are typically preceded by increased vocalisation (37).

We define a group that contains a pre-specified number of family units (n=20), reproductive pairs (n=10), and unpaired individuals (n=2). We stochastically determine family unit size to be between 3 and 8 individuals (38). We distribute family units (here including reproductive pairs within our definition) uniformly at random across 2D space, with the spatial location of individuals drawn from a normal distribution centred on the family centroid. As a simplifying assumption we assume individuals remain static—equivalent to a resting group or a foraging group over a short timescale.

We then define subgroup membership by constructing a social proximity network defining proximity as two individuals within a threshold distance. Each connected component in the proximity network represents a subgroup hyperedge. A second set of hyperedges connects individuals within the same family unit (including reproductive pairs). We can alternatively represent these family and subgroup networks as dyadic networks in which pairs of individuals are connected by edges if they are both in the same sub-group or family unit.

We then simulate the transmission of calling behaviour across the group. First, we define a threshold audible radius over which individuals can hear each other and use this to generate a network to indicate who can hear whom. We then select two individuals at random to be initial callers. We model the uptake of calling behaviour as the function of a dose-response curve, in which the "dose" is modified by the social relationship between two individuals. We assume the probability of an individual transitioning from a non-calling to a calling state is a function of the dose. In our network model, we calculate the probability of transition independently for each dyadic connection and take a corresponding draw from a Bernoulli distribution. We consider an individual to have started calling whenever at least one possible transmission event occurs. In our hypergraph model, we calculate the total "dose" an individual receives based on its hyperedge membership. We use this total dose to calculate the transition probability of an individual from non-calling to calling. The transmission model is equivalent to that described in (27). We model the behavioural contagion for 1000 timesteps (timesteps are arbitrary).

We then simulate departure decisions of individuals (i.e., transition from a present to a departed state) based only on the proportion of calling individuals within the audible radius. For these simulations, we set this threshold such that individuals only depart when 90% of others within this radius are calling. We record when each individual departs and whom it departs with.

Comparing the outcomes of our network and hypergraph models demonstrates that including a complex behavioural contagion specified on the hypergraph depiction of the social network leads to improved coordination in the departure of family units and sub-groups. When transmission depends on the higher-order social structure of the group then: a) the mean size of departing groups is larger (Fig. 2b); b) the departures of both family units (Fig. 2c) and sub-groups (Fig. 2d) are more coordinated; and c) fewer individuals are likely to remain (undeparted) at the end of the simulation (Fig. 2e) as compared to when transmission depends on dyadic interactions.

Our example highlights how incorporating higher-order social structure can help construct effective models of complex behavioural contagions within groups based on vocal communication. While you could model these same complex contagions using dyadic networks, it is more straightforward when conceptualised as a higher-order network. The results demonstrate that incorporating the higher-order structure found within groups can substantially alter the predictions made when quantifying the outcome of information transmission through groups using vocal communication. Given our hypergraph model predicts improved coordination in departure, especially among individuals that share higher-order social connections, it seems likely these types of social structures will be important in explaining partial- and full-consensus decisions during collective departures (as well as for other collective action problems). More generally, comparing the predicted outcomes of higher-order and dyadic network models with empirical data can be used to infer the best performing model and ultimately estimate the importance of higher-order social structures across different species. This can reveal how vocal communication contributes to complex behavioural contagions within groups.

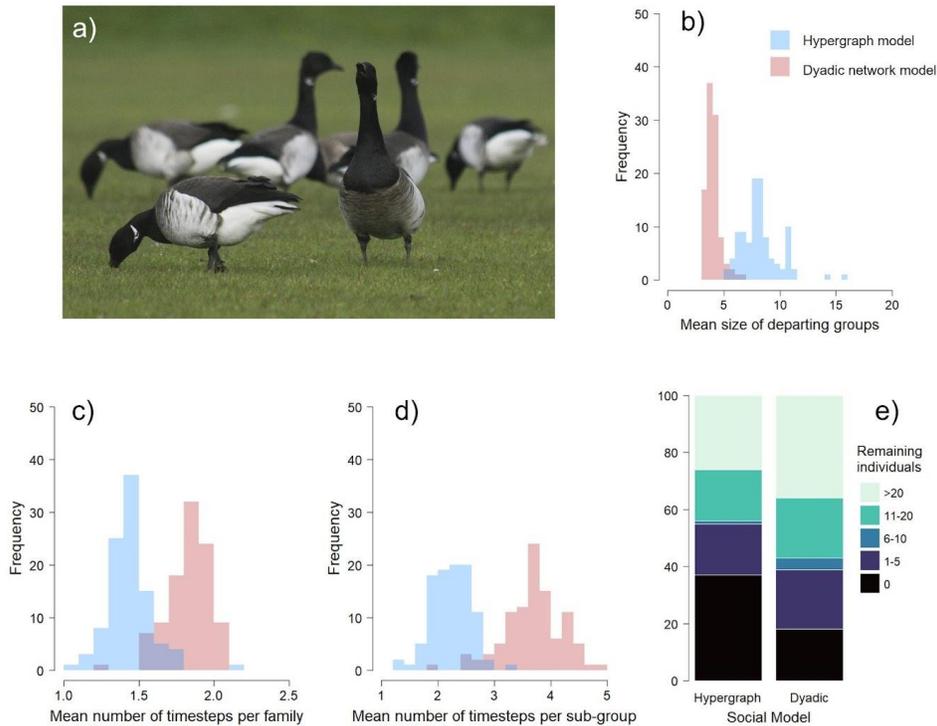

*Figure 2. An illustration of hypergraph and dyadic network models of the vocal coordination of collective departure. a) Light-bellied brent geese form fission-fusion social groups during non-breeding periods with foraging or roosting groups sub-structured by space and family ties. Hypergraph models of vocally-coordinated group departures inspired by this species predict b) larger departing sub-groups; c) greater coordination among family units in departure timing; d) greater coordination among spatially-defined units in departure time; and e) a greater frequency of all individuals being involved in partial consensus departure decisions than dyadic models.*

## 3.2 Case Study 2: Synchronisation in vocal signalling

Animal choruses are widespread (found in insects, anurans, fish, birds, and mammals) and impressive vocal displays characterised by high rates of signalling by many individuals. By their very nature, as opposed to duets, they are thus best described in terms of higher-order interactions because multiple signallers simultaneously advertise and could interact vocally with one or multiple nearby individuals. Additionally, choruses are dynamic, and individuals may be involved in multiple simplices at different points in time. Rich information is available to receivers from both individual and interactive communication displays. We illustrate how the dawn choruses of territory-holding songbirds may be explained by modelling their higher-order interactions using simplicial complexes; we showcase how many-body interactions could give rise to self-organised coordination within choruses.

Black-capped chickadee dawn vocal communication networks provide a good example to illustrate the types of higher-order interactions that may exist in territorial systems. Chickadees have a relatively unique social system among songbirds. In winter, several pairs and unpaired individuals form winter flocks with linear dominance hierarchies where males are dominant to females and older individuals dominate juveniles (39,40). In spring, pairs defend breeding territories within their winter flock home range (39). Female black-capped chickadees prefer higher-ranking males as both within and extra-pair mates (41–43). As such, information about social familiarity (flock membership) and social rank has been linked to vocal behaviour and fitness (44). Chickadees sing a pronounced dawn chorus in which all territorial males sing and choruses honestly signal both age and winter dominance rank (45,46).

Multi-microphone array recordings revealed that dawn chorus interactions are influenced by both winter dominance rank and flock membership, and that higher-order processes are features of these networks (47,48). Black-capped chickadees sing a simple two-note song 'fee-bee' that they can shift up and down a continuous frequency range (49). During vocal interactions or in response to song playback, males change their song frequency relative to their opponent (50), with frequency matching perceived as more aggressive than singing at a different frequency (51). Frequency matching occurs in dyads at dawn but can also include three or even four individuals (48).

Frequency matching patterns matched predictions based on social relationships, with more matching between males from different flocks or of similar dominance rank (47). Matching interactions extended from dyadic to triadic in a predictable way, beginning more often with two males from different flocks joined by a flockmate of one of them (48). In black-capped chickadees and other vocal territorial species, whether the synchrony in timing of the chorus or in timing of signals within the chorus is influenced by higher-order processes could be studied using the methods we outline here.

      As an example, we consider the configuration of territories from a population of another songbird species — ovenbirds *Seiurus aurocapilla* — for which we had GPS tracks of singing males used to map territories throughout the 2022 breeding season (Figure 3). We looked at the overlaps of buffered territory boundaries (Figure 3c) to construct an empirical higher-order structure that could exist in a vocal network of territorial ovenbirds. We used a 25m buffer beyond the territory boundary, in which we estimated most songs would be within the comfortable communication or recognition distances of neighbours as estimated previously for other songbird species (52) and as the likely area in which we would expect most higher-order interactions to occur. The higher-order structure (Figure 3d) suggests that simplices are likely features of territorial vocal communication networks and that some ovenbirds are involved in larger or more simplices than others. The simplices identified by the model match behavioural observations of multi-way vocal interactions documented during breeding activity surveys (J Foote *pers obs*).

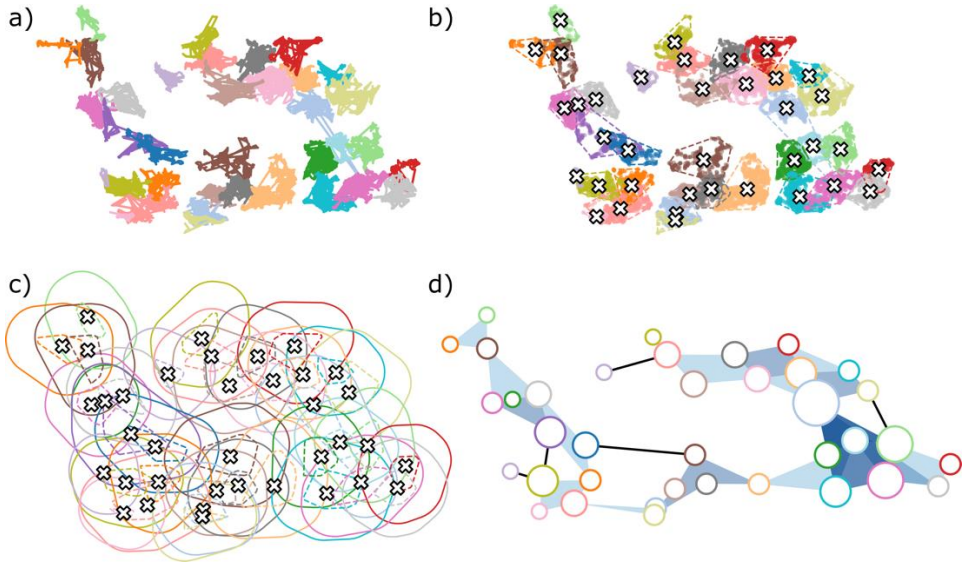

*Figure 3. Constructing a higher-order structure based on audibility range from empirical data of bird positions. a) Individual trajectories from GPS tracks of singing and foraging males where each colour represents the position in time of a different bird. b) The centroid of each trajectory is computed (white cross) together with the associated convex hull (dashed line) representing a proxy for the area covered by each bird. c) Each area is expanded by adding a buffer distance of 25m, in which we estimate most songs from within the territory would be recognized by neighbours representing the approximated hearing range. d) Finally, a simplicial complex is constructed from the many-body intersections within the buffered areas. The colour of the simplexes is proportional to the order; the size of the nodes is proportional to the number of attached simplices.*

We use the ovenbird position data (Fig. 3) as the backbone structure to run a higher-order generalisation of the classic Kuramoto model (53) proposed by Adhikari et al. 2023 (54). Coupled oscillators have been widely used to model emergent synchronisation patterns in various networked populations (55). We consider a hypergraph (Figure 4), where each bird is associated to a node and birds can interact by means of pairwise or three-way interactions (higher order interactions are possible but neglected for simplicity). Each node $i$ is associated to a natural frequency $\omega_i$ (randomly drawn from a normal distribution) and a state variable $\theta_i \in [0, 2\pi[$ representing the central "object" of the synchronisation dynamics, which could be, for example, the signal content (e.g., song frequency or type) or song timing (e.g., overlapping or alternating pattern of singing) to be matched upon an interaction. It is easier to think of this state variable, called the *phase* of the oscillator, as an angle in the unit circle. With this approach, we can easily quantify the degree of synchrony in our system by measuring how much our oscillators rotate together around the circle. After embedding the phases in the complex plane via the transformation $z = e^{i\theta}$, we can thus measure the level of synchrony as the average vector of these complex numbers:

$$z = \frac{1}{N} \sum_{i=1}^{N} e^{i\theta}$$

The magnitude $r = |z|$ represents the Kuramoto order parameter, also called phase coherence, that takes its maximum value $r = 1$ for perfectly synchronised phases. The equation of motions governing the dynamics of the $N$ nodes are:

$$d_t \theta_i = \omega_i + K_2 \sum_{j=1}^{N} a_{ij} sin(\theta_j - \theta_i) + K_3 \sum_{k,j=1}^{N} b_{ijk} sin(2\theta_k - \theta_j - \theta_i),$$

where the free parameters $K_2$ (dyadic) and $K_3$ (multi-body) are the coupling strengths associated to the interactions at different orders. Positive coupling will induce synchrony among the interacting individuals. In our case, these interactions are mediated by the 1- and 2-simplices encoded into the respective adjacency tensors: A represents the standard adjacency matrix — for 1-simplices —, with elements $a_{ij} = 1$ if nodes $i$ and $j$ share a 1-simplex (otherwise $a_{ij} = 0$). Similarly, the adjacency tensor B for 2-simplices will have elements $b_{ijk} = 1$ when nodes $i$, $j$, and $k$ participate in the same 2-simplex. We assume a simple sinusoidal function for the coupling, but the model naturally extends to interaction functions of any form.

The model suggests that synchronisation can emerge in neighbourhoods with higher-order interactions (Fig. 4). A periodic pattern of synchrony emerges (Fig. 4a) under moderate positive three-body coupling and mild divergent coupling in dyads. This pattern could emerge if chorusing individuals somewhat attend to their two closest neighbours simultaneously when singing but shift attention between neighbours over time as they move within their territory, leading to clusters of synchrony at different time points. If we increase the coupling strength of three-body interactions, the overall synchronisation of the whole chorus is reduced, and no local synchrony occurs (Fig. 4b). Finally, for positive coupling strengths for both dyadic and three-body interactions, we get synchronisation in clusters (Fig. 4c). Local patterns emerge, where neighbourhoods display phase-locking behaviour. This pattern likely matches nicely what we see in black-capped chickadee dawn choruses where individuals synchronise song frequency with individual neighbours in turn and also with two neighbours simultaneously (48), indicating similar behaviours could occur in ovenbirds and other species.

While these approaches remain untested in songbird territorial networks, pairwise phase coherence associated to volume oscillations as a function of their spatial distances has been shown in cicadas (56), with spatial synchrony also found to be patchy in this example. Similarly, in field crickets *Gryllus campestris*, with increased number of singing rivals, males were more likely to sing leading to moderate singing overlap but singing was inhibited by close proximity (57). Models of synchronisation have also been applied to frog choruses (58). Additionally, there is scope for higher-order vocal communication networks influencing synchrony across a wider diversity of species. For example, indris *(Indri indri)* use long-range songs to communicate between territorial groups (pairs and their offspring), and encounters in peripheral areas are mediated mostly through long-range vocalizations (59) meaning synchrony involving multi-body interactions could emerge where peripheral edges of multiple territories overlap. Further technological advances (e.g. in underwater recording) could extend research to aquatic systems such as territorial chorusing fish, for example the territorial defense and mate attraction calls of damselfish produce territorial calls used in both mate attraction and territorial defense (60).

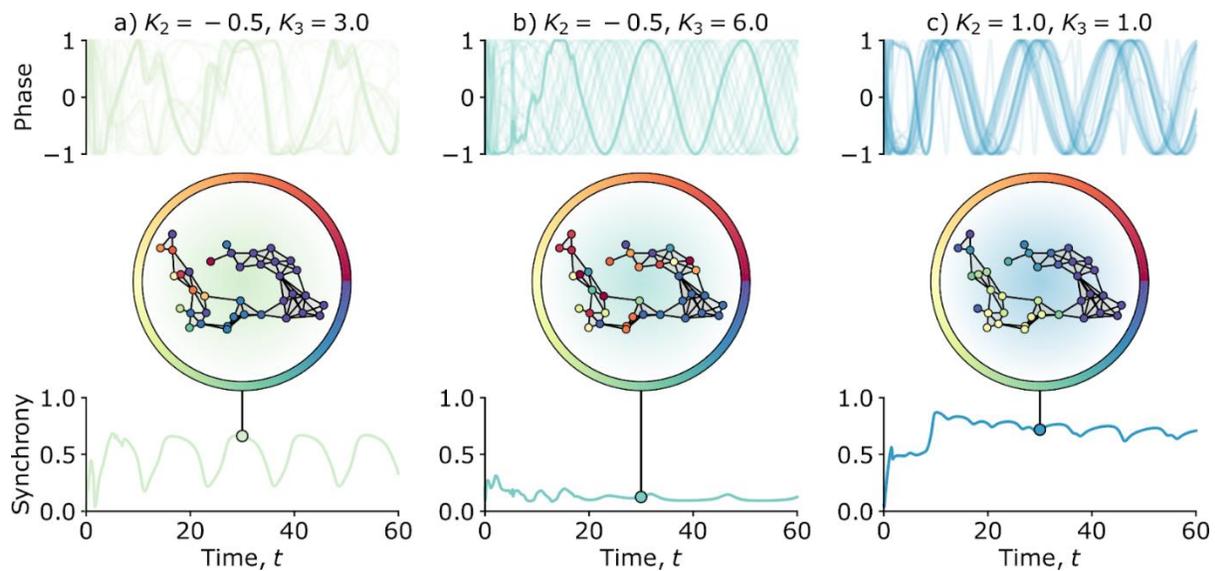

*Figure 4. Examples of a synchronisation problem mediated by higher-order interactions. Top panels report the temporal evolution of the phases associated to each node (sin($\theta$)), representing the song frequency. Bottom panels report the evolution of the associated order parameter — a proxy for the degree of synchrony of the system. For a visual aid, we also display the network together with the status of each oscillator at time t=30 by means of different colours picked from a circular colormap: nodes with similar colours will have similar phases. To illustrate the different behaviours that can be obtained through higher-order approaches, we consider three different scenarios of coupling strengths $K_2$ and $K_3$, where negative values indicate divergence and positive values convergence and the magnitude of coupling strength (e.g., how fast you try to keep up with neighbours or not) is variable. The interplay between these coupling parameters at different orders lead to different phenomena: a) a partially synchronised system with periodic signs of synchrony within the most connected nodes; b) an almost asynchronous system with no signs of coordination among the nodes; c) a system where clusters of nodes are synchronised.*

**Case Study 3: Long-term patterns of group coordination and culture**

Vocal communication signals are socially learned in many species (61). Socially learned signals often vary within and among social groups, through cultural fission and fusion (62). For example, male humpback whales *Megaptera novaeangliae* within particular populations in the Pacific Ocean conform to a particular song type, but due to aggregations of individuals from different populations at migratory stopovers, there is a striking pattern of directional cultural transmission on songs across the overall metapopulation (63,64). We can examine the fission-fusion dynamics of cultural groups using higher-order communication networks. Here, we focus on socially learned vocalizations—an important signal of cultural identity in many songbirds—to show how differences in who can hear whom will influence how different cultural groups form and are maintained.

White-crowned sparrows *Zonotrichia leucophrys* are a useful example to illustrate outcomes of higher-order interactions during the process of learning a social signal. Adult males typically produce one song type, and most males in an area share the same song type; a cultural group also known as a 'dialect'. Males from different areas produce different dialects (65). Cultural groups are thus separated geographically but not defined by geographic barriers – males can move between dialects (66). Males can memorize multiple song types from multiple tutors when young, overproduce these song types, and then selectively drop those that do not match the song of interacting males (66). Despite this selection for conformity, individuals are recognizable from their songs (67). There is subspecific variation in this song learning process, though, and it is thought that this variation could, in part, explain differences in the degree of geographic structure of dialects between resident and migratory subspecies (68).

One potential route to cultural fission is through detectability in the soundscape – if individuals cannot hear each other, they cannot influence each other's song. Decreasing coordination among individuals over spatial scales greater than an individual's audible range therefore leads to the emergence of different songs as individuals spread out or the population expands into new areas: group fission. Conversely, if individuals sing dissimilar songs within each other's audible range, they can learn from each other and songs can become more similar (within and across generations): group fusion.

As a potential explanatory model of these dynamics, we construct a higher-order network model in which interactions are determined by audible distance, $A$. In our model, individuals favour song variants within the boundaries of their cultural group norms, while maintaining sufficient distinctiveness to be individually recognizable. From just this simple scenario, we propose the following model:

Consider a population of individuals, $x_j$. We describe the geographic distance between the position, $x_j^p$, of any two individuals to be $GeoD(x_j, x_i) = GeoD(x_i, x_j) = |x_j^p - x_i^p|$. We describe their song, $x_j^S$, as a real valued number, such that we can define the distance between the songs of two individuals to be $SongD(x_j, x_i) = SongD(x_i, x_j) = |x_j^S - x_i^S|$. We then call $\epsilon$ the range of distinctiveness, such that if $SongD(x_i, x_j) \leq \epsilon$, then both individuals incur a fitness penalty due to lack of distinctiveness and therefore, if capable, alter their song.

We define a simplex $G\_x_j = \{x_i \mid i \neq j, GeoD(x_j, x_i) < A\}$. We then define $Med(G\_x_j)$ as the median value of the song of all individuals within $G\_x_j$

We can then define how $x_j^S$ changes over time by a two-step rule:

1) If there exists any $x_i$ such that $SongD(x_j, x_i) \leq \varepsilon$ then $\dot{x}_j^S = x_j^S \pm rj\varepsilon$, where $r$ is a random value drawn from the uniform interval $[-R, R]$. This moves the song away from its current value by a small random distance, pulled from a distribution tied to its ordinal position within the population. This decreases the likelihood that any two individuals with songs within the range of distinctiveness will simultaneously alter their songs to remain indistinct from each other after they both shift.

2) $\dot{x}_j^S = \begin{cases} x_j^S + \frac{Med(G\_x_j) - x_j^S}{f} & if\ Med(G\_x_j) \geq x_j^S \\ x_j^S - \left(\frac{x_j^S - Med(G\_x_j)}{f}\right) & if\ Med(G\_x_j) < x_j^S \end{cases}$ where $f$ is an integer ≥2 to reflect song flexibility over time. (nb. 1 indicates maximum flexibility, allowing individuals to converge to their audible community's median song). This moves each individual's song closer to the median within their audible range, $A$.

This model can either be interpreted to reflect an individual changing their own song (66), or an individual's offspring inheriting parental territory and learning/defining their own song (69).

We consider a discrete time simulation of this model under two scenarios. In the first scenario, individuals do not move and the only differences will arise as an outcome of changes in song, as influenced by others in a static set of simplices. In the second scenario, individuals move according to a standard flocking behaviour model (70). The simplices therefore change over time depending on movement-altered geographic distance altering who can hear whom, but otherwise independent of song.

From this simple model, we contrast the impact of static versus dynamic simplicial structures on population-wide emergent patterns in song (Fig 5). Individuals within the static spatial structure have more dissimilar songs (Fig 5b), consistent with the lack of interaction fostering divergence among subgroups akin to more geographic structure of dialects in resident white-crowned sparrow subspecies as compared to more migratory ones. However, a more nuanced story is revealed by examining the covariance between geographic and song distance over time in the two scenarios. While highly variant across scenarios, there are indications that geographically closer individuals have more dissimilar songs on average in the stationary scenario, while in the dynamic scenario, the mean scenario is for geographically close individuals to initially converge in song distance before then diverging to a more strongly negative correlation (Fig. 5c). Such a prediction has not been considered nor tested in this long-studied model species. This finding would be consistent with the initial formation of smaller groups with their own cultural identity and converging songs, but then fission-fusion dynamics in song-independent flocking behaviour causing these small cultural groups to encounter each other, intermix and form new small groups with different sets of individuals more rapidly than the convergence in song allows.

Our higher-order network model helps us characterise features that would be opaque in a dyadic framework. For example, one native output is a prediction of fission-fusion dynamics over time and space. An easily measurable feature of the emergent dynamics is the overlap in subsimplices shared by groups, i.e., "sub-simplicial density". Greater numbers of shared subsimplices would suggest maintenance or emergence (i.e., fusion) of a single cultural group, whereas regions of sparse sub-simplicial density between two clusters of greater density would suggest fission. In this way, we can predict the emergence or loss of cultural groups from a landscape. This model is easily extendable to other songbirds which vary in their song learning process along multiple dimensions (71). For instance, we could also predict the song-trajectory of individuals from open ended learners over their lifespans, depending on their geographic location and the influence of audible cultural groups. The group's cultural norms themselves will shift over time as individuals change their songs and move into and out of audible range. Extending this model to bias flocking behaviour towards similar song could then yield complex bidirectional feedback between song type and movement, generating valuable predictions about the relationships between culture formation and movement ecology.

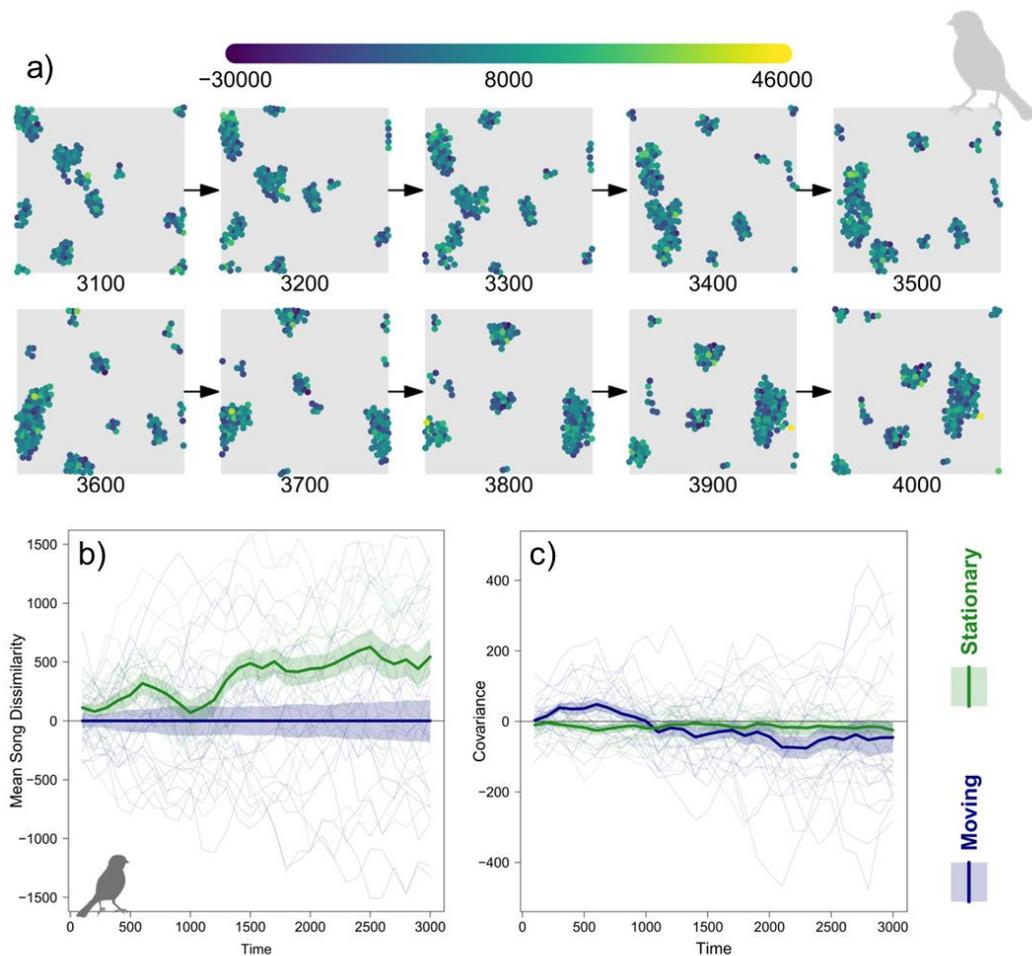

*Figure 5. An example of our Case Study 3 model in action. a) An illustration of a single model run of the "moving" condition in which individuals follow basic rules of collective action. Points are coloured according to their song values. Only the final part of the simulation is illustrated but the full simulation can be watched in the Supplementary Video. b) An illustration of how song dissimilarity increases faster for the "stationary"*

*than the "moving" condition for the selected parameter set. All values are scaled to the mean of the "moving" condition for illustrative purposes. c) An illustration of how the covariance between song dissimilarity and distance between individuals changes over time under the "moving" and "stationary" conditions. For panels b) and c) Time is the time since individual song values were free to change (i.e., from time step 1000). The thick lines show the mean song dissimilarity over time for 20 simulations runs and the shaded area the standard errors around the mean. Semi-transparent lines illustrate the mean song dissimilarity in each simulation run. Parameters used for these simulations were $R = 5, A = 100, f = 500$ and $\varepsilon = 50$.*

## 4. Toolkit for higher-order networks

Accessible software tools are not as well developed for higher-order network approaches as they are for dyadic networks, although this is changing rapidly. There are recent broad overviews of higher-order methods (12,13) and Silk et al. (27) provide an overview of the main tools available to ecologists, so here we focus on the most relevant tools available. The XGI *Python* library (72) provides a comprehensive toolkit for working with higher-order networks. It includes tools for calculating descriptive measures, some basic generative models, and visualisation tools alongside flexible core data structures for handling, converting, filtering, and storing hypergraphs and simplicial complexes. Alternatives in Python also include HypergraphX (73), while scikit-TDA represents a more general collection of libraries for Topological Data Analysis (74). In *R*, the tdaverse (https://github.com/tdaverse/tdaverse) provides tools for working with simplicial complexes, including plotting of Vietoris-Rips and Čech complexes while the packages HyperG (75) and rhype (76) provide various algorithms for descriptive measures, basic generative models and plotting. Basic tools for working with hypergraphs and simplicial complexes are also available in *Julia* (e.g. SimpleHypergraphs.jl: (77); Simplicial.jl: https://github.com/nebneuron/Simplicial.jl).

## 5. Key considerations and challenges

It is important to discuss the general applicability of the proposed framework. An important caveat is that higher-order network approaches only benefit cases to which the theory applies well; just because something can be described as a hypergraph or simplicial complex, does not mean that it is always convenient to adopt such representation, nor that the higher-order dynamics are ecologically or evolutionarily meaningful. Some theoretical constructs may not be of practical use in real-world systems if assumptions central to the model are not realistically satisfied by any real-world scenario. There will also be cases where, even if the observed system naturally fits within the domain of many-body interactions, pairwise approaches already provide enough complexity to correctly capture the essential underlying mechanisms. For example, recent results have shown that dynamical systems defined on hypergraphs (at the node level) can be effectively reduced to dynamics on dyadic networks if the many-body interactions enter exclusively through linear functions of the states of the nodes (78). Other recent advances have also shown that, in some cases, it is possible to "lower" the order of some hyperedges without affecting the dynamical outcome — reducing the overall complexity of the model (79). In general, Occam's razor remains a core principle: if a pairwise model is sufficient to explain the observed dynamics, there is no need to go higher with the order of the interactions. Higher-order models could indeed present further computational challenges that one might want to avoid if not strictly necessary. Consequently, methods that could provide meaningful insight, and satisfy the assumptions that make their interpretation valid, are limited by tractability. This can be particularly common as the size of simplices or hyperedges increases. In these cases, the application of higher-order network methods is ill advised unless limiting them to hyperedges or simplices of a particular size (or smaller) can be justified. As with all research, these limitations do not represent firm endpoints, but opportunities for cutting edge methodological research to extend the boundaries of capability.

## 5.1 Data requirements

A key challenge to applying higher-order network approaches to study animal vocal communication will be collecting appropriate data, something that is only recently possible with the advancement of acoustic and biologging technology. We envision an ideal data collection protocol to gather simultaneously both movement and vocalization data of tracked individuals, for example, using acoustic location systems (80) or animal-borne microphones (81) or accelerometers (82). In general, it will be important to obtain longitudinal data, akin to recent efforts in human social networks (83), whose spatial and temporal extent goes beyond more traditional data collection procedures.

Quantifying the distance of communication in territorial networks will also be important to effectively model higher-order communication networks. We need to quantify not only how many individuals a receiver can hear, but also the distance at which information can be extracted (84). These distances can vary among species and with background noise levels (85). A further challenge is that for species with a single vocalisation type that is relatively invariable, it can be difficult to discern several independent signalers from dyadic or higher-order interactions. However, advances in autonomous acoustic location systems (86) could record networks of territory holders where position information can alleviate these issues.

**5.2 Model fitting versus prediction**

Following the discussion above, a key challenge in applying higher-order network models to communication networks will be in finding effective ways to: a) parameterise them to fit empirical datasets; and b) make suitably cautious predictions about how the system might respond to demographic or environmental change. For the former, choosing the correct outcome measures to compare between the empirical data and model predictions will be important. A second important point is that just because a higher-order model makes predictions that fit an empirical dataset does not necessarily mean we have identified the correct underlying mechanism. In fact, in absence of a ground truth — that is how nature operates — model selection becomes particularly crucial, especially against models that only incorporate dyadic network structures, to mitigate the risk of overfitting (the same argument also applies when adding in individual heterogeneity, temporal dynamics, etc.). For the latter, a potential strength of using these approaches is to forecast either ecological or evolutionary responses to scenarios such as reduced population density, habitat fragmentation or anthropogenic noise. However, these predictions must be made carefully as it is challenging to accurately parameterise how individual behaviour may change in response to these factors, and this would alter the higher-order network structure. For example, if individuals respond to anthropogenic noise by moving more around their territory or targeting vocal communication at specific individuals, then predictions that did not incorporate these changes would likely be incorrect. Consequently, we advocate that when making predictions related to higher-order communication networks, researchers are very clear about their assumptions and any caveats to interpretation.

## 6. Concluding remarks and future directions

As demonstrated by our case studies, higher-order networks allow us to address questions that earlier models could not easily interrogate on behavioural, ecological, and even evolutionary scales. This means that not only already-posed questions can be revisited and addressed anew, but also future studies may benefit by framing hypotheses and gathering empirical data to anticipate the parameterization and/or validation of higher-order models.

### 6.1 The value of interdisciplinarity

Animal vocal communication network research offers an exciting opportunity to shape the direction of theoretical and methodological advancements in network science across boundaries of traditional research disciplines. Current advances in higher-order network modelling have been predominantly shaped by applications to sociological and engineering questions, and the intrinsic intuition of network scientists and complexity theorists. It is tantalising to imagine what new capabilities can be developed once interdisciplinary teams work together to envision their role in biological discovery. Through these collaborations "if only this concrete next concept were measurable, quantifiable, or computationally tractable to analyse" becomes a call to action, shaping basic methodological research. This is an opportunity for true bi-directional synthesis among biologists and network scientists that can extend the scope of both disciplines and answer questions together that would otherwise have been out of reach.

Two key examples of where animal vocal communication networks can inspire the development of new theoretical models are in the incorporation of individual heterogeneity and the development of new models for the temporal dynamics of higher-order networks (87). Individuals may vary in how they receive and respond to signals, as well as their loudness of vocal communication. Developing approaches to incorporate this heterogeneity into theoretical models of higher-order social and communication networks could have important implications for general understanding of their dynamics. Similarly, developing computationally-tractable methods to deal with new model challenges brought by temporally-dynamic higher-order networks is a key area of methodological development that could be influenced by empirical challenges in animal communication. For example, a truly temporally explicit extension of Case Study 2 that accounts for the movements of each individual would: a) require highly sophisticated data collection that tracked the location of every vocalisation of each individual (e.g. a time-synchronised recorder array; (88)); b) increase the complexity of the analysis pipeline to correctly identify the type or frequency of song while accounting for changes in background noise; c) induce further computational challenges to identify the correct temporal scales for modelling the dynamical process.

## 6.2 Conclusion

Researchers studying animal vocal communication have long recognised that these interactions are not dyadic, and their non-dyadic nature has important ecological and evolutionary consequences. We highlight how developments in modelling higher-order networks provide tools to explicitly account for these non-dyadic interactions. Our case studies demonstrate how higher order approaches can transform our understanding across diverse social contexts and help tackle new research questions. We also show the potential for animal communication networks to inspire and test new theoretical models in network science. Overall, we hope this article provides the motivation and tools to develop and test new models of higher-order communication networks in the wild.

## Acknowledgements

We thank two anonymous reviews for helpful comments on an initial draft of this manuscript.

**Funding**

II was partially supported by the James S. McDonnell Foundation 21$^{st}$ Century Science Initiative Understanding Dynamic and Multi-scale Systems - Postdoctoral Fellowship Award. JRF received funding from the National Science and Engineering Research Council of Canada Discovery Grant program. MJS received funding from the European Union's Horizon 2020 research and innovation programme under the Marie Sklodowska-Curie grant agreement No. 101023948 and from a Royal Society University Research Fellowship URF\R1\221800.